\documentclass[twocolumn,preprintnumbers,aps,prb,superscriptaddress]{revtex4}

\usepackage{graphicx,times,epsfig,color}
\usepackage{float}
\usepackage{amsmath} 
\usepackage{xfrac}
\usepackage{braket}
\usepackage{amssymb}    
\usepackage{graphics} 

\usepackage{comment}

\begin{document}

\def\salto{\vskip 1cm}
\def\lgr{\langle\langle}
\def\rgr{\rangle\rangle}

\title{Spin-1 two-impurity Kondo problem on a lattice}
%
\author{A. Allerdt}
\affiliation{Department of Physics, Northeastern University, Boston, Massachusetts 02115, USA}
\author{R. \v{Z}itko}
\affiliation{Jo\v{z}ef Stefan Institute, Jamova 39, SI-1000 Ljubljana, Slovenia \\
Faculty of Mathematics and Physics, University of Ljubljana, Jadranska 19, SI-1000 Ljubljana, Slovenia}
\author{A. E. Feiguin}
\affiliation{Department of Physics, Northeastern University, Boston, Massachusetts 02115, USA}

\begin{abstract}
We present an extensive study of the two-impurity Kondo problem for
spin-1 adatoms on square lattice using an exact canonical transformation to map the problem onto an effective one-dimensional system that can be numerically solved using the density matrix renormalization group method. 
We provide a simple intuitive picture and identify the different
regimes, depending on the distance between the two impurities, Kondo coupling $J_K$, longitudinal anisotropy $D$, and transverse anisotropy $E$.
In the isotropic case, two impurities on opposite(same) sublattices have a singlet(triplet) ground state. However, the energy difference between the triplet ground state and the singlet excited state is very small and we expect 
an effectively four-fold degenerate ground state, {\it i.e.}, two decoupled
impurities.
For large enough $J_K$ the impurities are practically uncorrelated forming two independent underscreened states with the conduction electrons, a clear non-perturbative effect. 
When the impurities are entangled in an RKKY-like state, Kondo correlations persists and the two effects coexist: the impurities are underscreened, and the
dangling spin-$1/2$ degrees of freedom are responsible for the
inter-impurity entanglement. We analyze the effects of magnetic anisotropy in the development of quasi-classical correlations.
\end{abstract}

\maketitle
\section{Introduction}
Magnetic adatoms can be used to tailor the properties of substrates\cite{Donati2014,Hu2012,Drost2017} and to build nanostructures that could pave the way to magnetic devices with novel functionality including qubits and magnetic memories\cite{Enders2010,Leuenberger.Quantum_computing,Ardavan.Will_spin_relaxation,Karlewski.Magnetic_adatoms_memory,Serrate2010,Donati2016,Natterer2017}. 
Exciting pioneering examples include ‘quantum corrals’ formed by Fe atoms on Cu(111) \cite{Crommie1993}, quantum wires on Au(111) surfaces\cite{Hasegawa1993}, and more recently, magnetic nanostructures\cite{Loth2012}, one dimensional atomic chains 
\cite{lagoute2007,Neel2011}, and atomic dimers \cite{Chen1999}.
Depending on their strength and directionality \cite{Otte2008,Zhou2010,Pruser2011,Pruser2014}, quantum-mechanical effects could lead to the self-assembly of one-dimensional Co chains and Fe superlattices\cite{Manoharan2000,Khajetoorians2012}.
These magnetic nanostructures can also serve as a platform to simulate magnetic quantum matter in the same spirit as cold atomic systems\cite{Wiesendanger2009,Wiesendanger.Atom_by_atom,Spinelli2015,Bercioux2017}.

This research field is rapidly developing, but creating new
technologies based on magnetic nanostructures requires a deeper
knowledge of the interplay between the different degrees of freedom
and energy scales in the problem. These vary widely depending on the
atomic species and the substrates. Some systems have particularly
distinguished properties. For instance, scanning tunneling
microscopy (STM) studies found large magnetic anisotropies for single
manganese and iron atoms on copper nitride.\cite{Hirjibehedin.Large_magnetic_anisotropy} 
In addition, the anisotropy can be controlled through a coupling of the spin with a conductive electrode\cite{Oberg.Control_single_spin}. 
More recently, a holmium atom on a platinum (111) surface was found to
have a very large total angular momentum of $J=8$.\cite{Karlewski.Magnetic_adatoms_memory}.

Most of the theoretical understanding relies on the theory of the
single-impurity problem\cite{HewsonBook} and indirect exchange
(Ruderman-Kittel-Kasuya-Yosida --
RKKY)\cite{RKKY1,RKKY2,RKKY3,Doniach1977}. The latter, in particular, is based on second-order perturbation theory and leaves aside many important puzzles that arise when correlations are taken into account non-perturbatively. 
For instance, a remarkable result in early numerical renormalization
group (NRG) studies of the two-impurity problem
\cite{Jones1988,Jones1989,Zarand2006} with spins $1/2$ indicated the
existence of a non-Fermi-liquid (NFL) fixed point. A different NFL fixed point was found in the three-impurity problem in the presence of magnetic frustration\cite{Ingersent2005}.

The two-impurity problem has been theoretically studied using a range of techniques,
both analytical and numerical. Introduced in
1981, it was first analyzed using perturbative scaling ideas \cite{Jayaprakash.Two_Kondo}.
Our understanding continued to evolve through the application of new
methods
\cite{fye1987,jones1987,jones1989prb,affleck1992,sire1993,affleck1995,gan1995prl,silva1996,georges1999,izumida2000,lopez2002,aguado2003,simon2005}
and the development of state of the art experimental setups
\cite{Chen1999,jeong2001,craig2004,wahl2007}. 
In most studies, the impurities
are described either by spin-$\frac{1}{2}$ Kondo model or single-orbital
Anderson model\cite{HewsonBook}. However, typical experiments with
adatoms on substrates involve transition-metal atoms with high spin such as iron, cobalt, or manganese.  

The minimal model to tackle this problem is the two-impurity Kondo Hamiltonian:
\begin{equation}
H =  H_{\rm band} + J_K \left( \vec{S}_1 \cdot \vec{s}_{\rm r_1} + \vec{S}_2 \cdot \vec{s}_{\rm r_2} \right).
\label{Hamiltonian}
\end{equation}
where $H_{\rm band}$ is the lattice Hamiltonian for non-interacting
electrons, $\vec{S}_i$ are the quantum-mechanical impurity spin operators, and $\vec{s}_{r_i}$ represents the
conduction electron's spin at the impurity's coordinate $r_i$, for impurities $i=1,2$.

In this work we focus on spin-1 impurities. 
A point-like spin-1 impurity coupled to a single conduction electron
channel will be "underscreened", {\it i.e.}, the ground state will
have a spin-$1/2$ residual moment \cite{Mattis.Symmetry_ground_state,
Nozieres1980}. In other words, a single electron channel can screen at
most one-half unit of spin. This problem has been extensively studied
and was exactly solved with the Bethe ansatz for the case of a linear
dispersion
\cite{Fateev1981,Fateev1981a,Furuya1982,andrei1983solution}, and
numerically using the numerical renormalization group (NRG) 
\cite{Cragg1979,Lloyd1980,Mehta2005,Koller2005,Zitko.Properties_of_magnetic_impurities}
for generic situation. The underscreened nature of the ground state
with two-fold degeneracy leads to singular thermodynamic behavior
characterized by a singular-Fermi-liquid fixed point\cite{Coleman2003,Mehta2005,Logan2014}. 
However, most physical examples have a natural magnetic anisotropy that arises from the local spin-orbit coupling that can be quite large\cite{Ujsaghy1999,Szunyogh2006,Etz2008}. 
Anisotropy is often modeled by adding a term:
\begin{equation}
\label{anisotropy}
H_A = DS_z^2 + E(S_x^2-S_y^2).
\end{equation}
Such interactions emerge as an effective spin Hamiltonian of an orbitally
degenerate level from the combination of spin-orbit coupling and
crystal-field splitting according to the point-group symmetry.  For
$S=1$, the physics strongly depends on the 
longitudinal anisotropy $D$, since it splits the three-fold degeneracy of the free
moment into a non-degenerate state $\left( S_z=0\right)$ and a pair of
states$\left( S_z=\pm1\right)$. 
The transverse anisotropy term $E$ is smaller than $D$, and zero if the point
group has a $C_N$ axis with $n \geq 3$. When non-zero, it splits the
$S_z=\pm 1$ pair. The transverse anisotropy is particularly important in the
context of spin-state life-times and decoherence \cite{delgado2017}.
The magnetic anisotropy hence determines the effective impurity
degrees of freedom and the possible physical mechanisms for their
quenching at low temperatures.
For example, STM experiments with cobalt atoms ($S=\sfrac{3}{2}$) on
Cu$_2$N find Kondo effect even in the
presence of a hard axis (positive $D$) anisotropy\cite{Otte.Role_of_anisotropy}. The Kondo effect is
observed if the ground state levels are degenerate and connected by
$\Delta m=1$. This occurs only for half-integer spin
$S\geq\sfrac{3}{2}$ when the anisotropy $D$ is positive. The Kondo effect
is absent for iron ($S=2$) and manganese ($S=\sfrac{5}{2}$) that both
have negative anisotropy.

%

The effects of anisotropy in the single-impurity problem are discussed
in detail in Ref.~\onlinecite{Zitko.Properties_of_magnetic_impurities}
using the NRG and a flat density of states. Although this problem
exhibits universal physics that is essentially independent of the band
structure
\cite{nrg_wilson,Andrei1980diagonalization,andrei1983solution,tsvelik1983I,tsvelik1983II,nrg_bulla},
the lattice plays an important role in mediating the inter-impurity coupling
and leads to the competition between different
effects\cite{Schwabe2012,Allerdt.Kondo,Mitchell2015,Schwabe2015}. The
complexity in the two-impurity case arises from competing energy
scales between the single-impurity screening and the RKKY physics. In
addition, for high spin the magnetic anisotropy also plays an important role
selecting the relevant degrees of freedom that participate in these
screening processes.  In order to study this problem on a specific
lattice, we use a recently developed computational technique
\cite{Allerdt.Kondo} that allows us to exactly solve this problem on
large lattices with the aid of the density matrix renormalization
group method (DMRG)\cite{White1992,White1993,schollwock2005density}.
Our results are exact and provide a deeper understanding of the
physics beyond perturbative ideas, accounting for all the
many-body effects.

The paper is organized as follows: In Section \ref{Strong} we present
a qualitative overview of the problem from the perspective of the
strong-coupling limit. This provides valuable intuition that will help
us interpret the numerical results. Section \ref{Method} describes
the exact mapping of the model onto an effective one-dimensional
problem amenable to DMRG calculations. In Section \ref{Results} we
present our results for both the isotropic and the anisotropic cases.
We finally close with a summary and conclusions.

\section{Qualitative picture}\label{Strong}

\subsection{Single spin-1 impurity}\label{Single}

\subsubsection{Underscreening}

\begin{figure}
\centering
\includegraphics[width=0.35\textwidth]{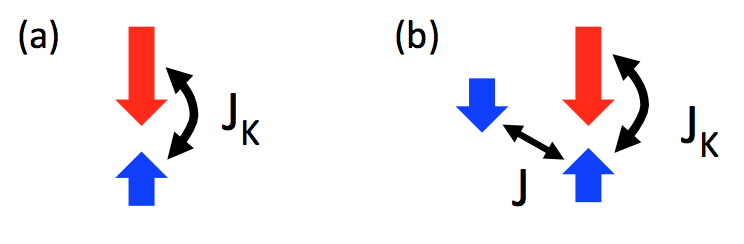}
\caption{Toy model representing the strong-coupling limit. Large red
arrows represent $S=1$ impurity spins, while small blue arrows
correspond to conduction-band electron spins. (a) A single conduction
spin can only partially screen the impurity. (b) A second conduction
spin interacts with the first through an antiferromagnetic exchange
$J$. The impurity spin and the second spin favor ferromagnetic alignment. }
\label{fig:single_model}
\end{figure}

\begin{figure}
\centering
\includegraphics[width=0.2\textwidth]{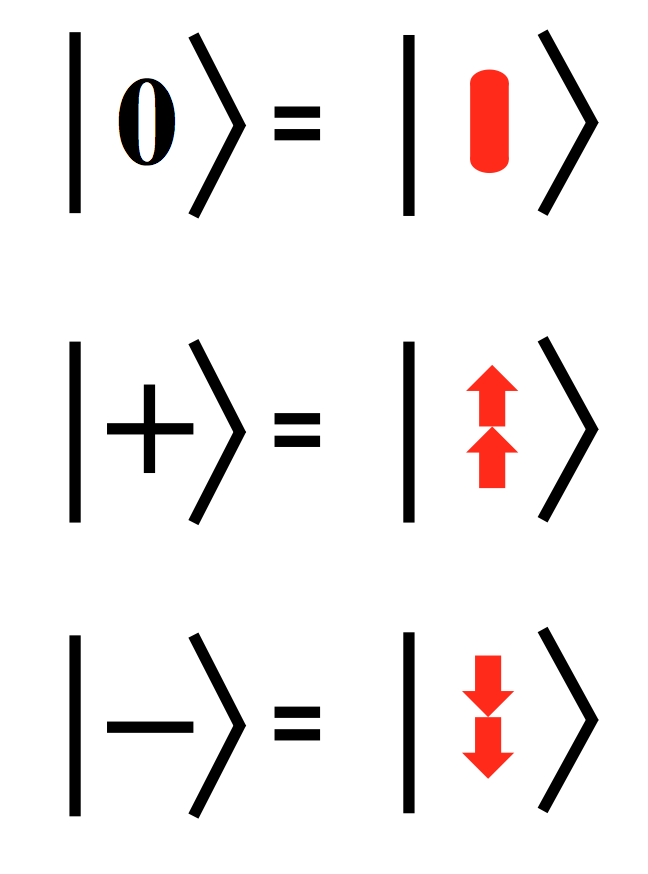}
\caption{Graphical representation of three basis states for a single
spin $S=1$ in terms of two spins $S=1/2$. The $|0\rangle$ state is a
triplet with total $S^z=0$, while the two other states are fully polarized.
}
\label{fig:single_spin}
\end{figure}

\begin{figure}
\centering
\includegraphics[width=0.35\textwidth]{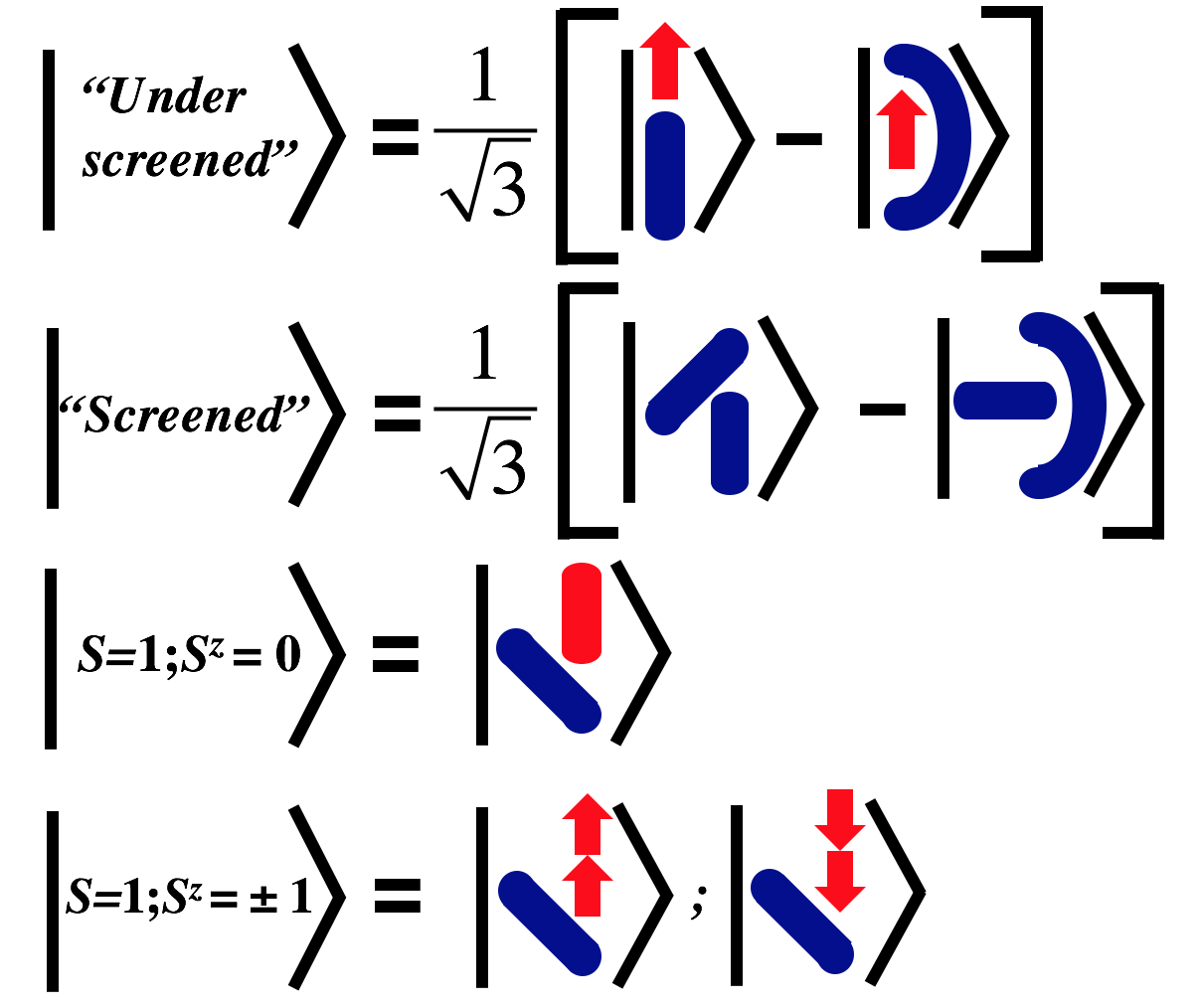}
\caption{(color online) Valence bond representation of the single site
``Underscreened'' spin-doublet state $|m_s=1/2\rangle$ with one conduction-band
electron spin, and of the possible ground states after adding a second
conduction spin: ``Screened'' singlet, and $S_z=0,\pm1$ triplet states.
The spin-1 impurity is represented as two spins $S=1/2$. The dangling unscreened spins are shown explicitly and the thick blue(red) lines represent singlets(triplets) between spins-1/2.
}
\label{fig:screening}
\end{figure}

We first propose an intuitive picture to understand the essential
physics of the underscreened Kondo problem, focusing on the $S=1$
single-impurity case, and use a very simple effective model for
illustrative purposes. It has been demonstrated in
Ref.~\onlinecite{Yang2017} that in the Kondo problem, only one conduction electron is responsible for the screening. This picture is quite universal. In the weak-coupling limit, one can imagine that this electron is located at the Fermi level \cite{schwabe2012competition}, while in the strong-coupling limit, it corresponds to the localized orbital directly in contact with the impurity spin. In either case, the remaining conduction electrons form a completely disentangled Fermi sea, and do not play a role in the physics. Therefore, this can be reduced to a problem of an impurity spin (partially) screened by a single conduction electron. 
For a spin-$1/2$ impurity, the ground state is a singlet between both
spins while for a spin-1 impurity, one needs to also account for the
magnetic anisotropy. We consider the impurity and conduction spins
interacting via a Heisenberg exchange $J_K$, and include the
anisotropy as in Eq.(\ref{anisotropy}), see
Fig.~\ref{fig:single_model}(a). Defining states
$|\uparrow\rangle,|\downarrow\rangle$ for the conduction spin, and
$|+\rangle,|0\rangle,|-\rangle$ for the impurity spin, one can readily
obtain the ground state as a function of $D$ (for now $E=0$).
The Hamiltonian can be diagonalized in blocks, and we focus on the subspaces with total $S^z=m_s=\pm 1/2$. The $2 \times 2$ Hamiltonian matrix is:
\begin{equation}
H(m_s=\pm1/2) = 
\begin{pmatrix}
0 & \frac{J_K}{\sqrt{2}} \\
\frac{J_K}{\sqrt{2}} & -\frac{J_K}{2}+D
\end{pmatrix} \quad .
\label{h1}
\end{equation}
For $D=0$ we find two degenerate ground states:
\begin{eqnarray} 
\ket{m_s=1/2} & \equiv & a \ket{0 \uparrow} - b\ket{+ \downarrow}; \nonumber \\
\ket{m_s=-1/2} & \equiv & a\ket{0 \downarrow} - b\ket{- \uparrow}, 
\label{ms}
\end{eqnarray}
with $a=\sqrt{1/3}$ and $b=\sqrt{2/3}$. This degeneracy is a
manifestation of underscreening: if one assumes that the spin-1 is
actually comprised of two spins $1/2$ (see
Fig.~\ref{fig:single_spin}), the conduction electron will screen only
one of them. This leaves a ``dangling'' spin-$1/2$ that can point in
either direction. To make this analogy more explicit, the
$S=1$ states are written in terms of two $S=1/2$ spins projected onto the $S=1$
manifold:
\begin{eqnarray}
|+\rangle & \equiv &  |\uparrow\uparrow\rangle \nonumber \\
|-\rangle & \equiv &  |\downarrow\downarrow\rangle \\
|0\rangle & \equiv &  \frac{1}{\sqrt{2}}\left(|\uparrow\downarrow\rangle +|\downarrow\uparrow\rangle \right) \nonumber 
\end{eqnarray}
We can readily re-write the eigenstate $|m_s=1/2\rangle$ as:
\begin{eqnarray}
|m_s=1/2\rangle & = & \frac{1}{\sqrt{6}}\left(|\uparrow\downarrow\rangle|\uparrow\rangle - |\uparrow\uparrow\rangle|\downarrow\rangle\right) + \nonumber \\
& + & \frac{1}{\sqrt{6}}\left(|\downarrow\uparrow\rangle|\uparrow\rangle - |\uparrow\uparrow\rangle|\downarrow\rangle\right).
\end{eqnarray}
The two terms in the above sum correspond to two singlets: one between
the conduction spin and the first spin, and the other between the
conduction spin and the second spin. This can be interpreted as a
resonating valence bond (RVB) state with an extra spin always pointing up (see 
$|\text{Underscreened}\rangle$ in Fig.~\ref{fig:screening} for a
graphical representation of this state).
It is easy to show that the entanglement entropy for the impurity is $S=- \frac{1}{3} \log{(\frac{1}{3})} - \frac{2}{3}\log{\frac{2}{3}}$.

A similar analysis applies to the $|m_s=-1/2\rangle$ state. 
The degeneracy of the $|m_s=\pm 1/2\rangle$ states leads to the
so-called singular Fermi liquid behavior that emerges when trying to
restore the time-reversal symmetry \cite{Coleman2003}. This can be
broken ``by hand'' by choosing a particular value of $S^z$. However, the symmetrization of the ground state implies that the self-energy of the problem has to be obtained from two distinct self-energies \cite{Logan2014}, yielding peculiar low-temperature effects. As soon as an infinitesimal magnetic field is applied, the system picks a unique ground state, and Fermi liquid behavior is restored.

The effect of the longitudinal anisotropy $D$ is to change the
character of the ground states to more ``classical'' or ``Ising-like''
states. $D \gg J_K $ implies $a \gg
b$, and the impurity spin has zero total moment. On the other hand, $D
\ll -J_K$ leads to $b \gg
a$, and the impurity spin is fully polarized in either direction. In
both limits, the impurity and the conduction spins are disentangled
and the screening is lost, but the two-fold degeneracy of the ground
state persists.  

We now analyze the effects of a transverse anisotropy parametrized by $E$ in Eq.~(\ref{anisotropy}). This term can be rewritten as
\begin{equation}
V_E=\frac{E}{2}\left[(S^+)^2 + (S^-)^2\right].
\label{transverse}
\end{equation}
 In an isolated quantum spin, this perturbation would mix 
 the $m=\pm 1$ states. However, this is no longer
 true in our ``underscreened Kondo'' problem. Let us first consider
 the two degenerate ground states in Eq.~(\ref{ms}), and apply
 perturbation theory. The first contribution comes from the
 second-order correction. It is diagonal and has the same sign for
 both $m_s$ states. This corresponds merely to a constant shift in the energy levels with no splitting. Another way to see this is by applying $V_E$ to the states directly: $V_E|m_s=1/2\rangle = -b/2|- \downarrow \rangle$, and $V_E|m_s=-1/2\rangle = -b/2|+ \uparrow \rangle$. 
Therefore, $V_E$ acts on separate subspaces and does not mix them.
Hence, we conclude that the transverse anisotropy does not modify the general picture described above.
The general form of the two degenerate states will now be
\begin{eqnarray}
|1\rangle & = & \alpha |0\uparrow\rangle +\beta|+\downarrow\rangle+\gamma|-\downarrow\rangle, \nonumber \\
|2\rangle & = & \alpha |0\downarrow\rangle +\beta|-\uparrow\rangle+\gamma|+\uparrow\rangle.
\label{states12}
\end{eqnarray}
Clearly, for $E=0$ we obtain $\alpha=a$, $\beta=b$, $\gamma=0$ and we recover the original $|m_s\rangle$ states.

\subsubsection{Residual interactions}

We should point out that strictly speaking the previous discussion
does not correspond to the strong-coupling limit of the problem. This is studied in Ref.~\onlinecite{Zitko.Properties_of_magnetic_impurities} and leads to XXZ anisotropic exchange constants. In addition, one has to account for the rapidly growing anisotropy (in the renormalization group sense) and a residual ferromagnetic coupling between the magnetic impurity and the conduction electrons. 
We follow Nozi\`eres and Blandin's simple arguments \cite{Nozieres1980}: an additional electron would want to increase its kinetic energy by hopping onto the site connected to the impurity. Due to Pauli's exclusion principle, it could only do that if it had the opposite spin orientation, which means the same spin orientation as the impurity spin. Using conventional perturbation theory arguments, this translates into an effective residual ferromagnetic coupling between the impurity spin and the remaining conduction spins. 

\begin{figure}
\centering \includegraphics[scale=0.55]{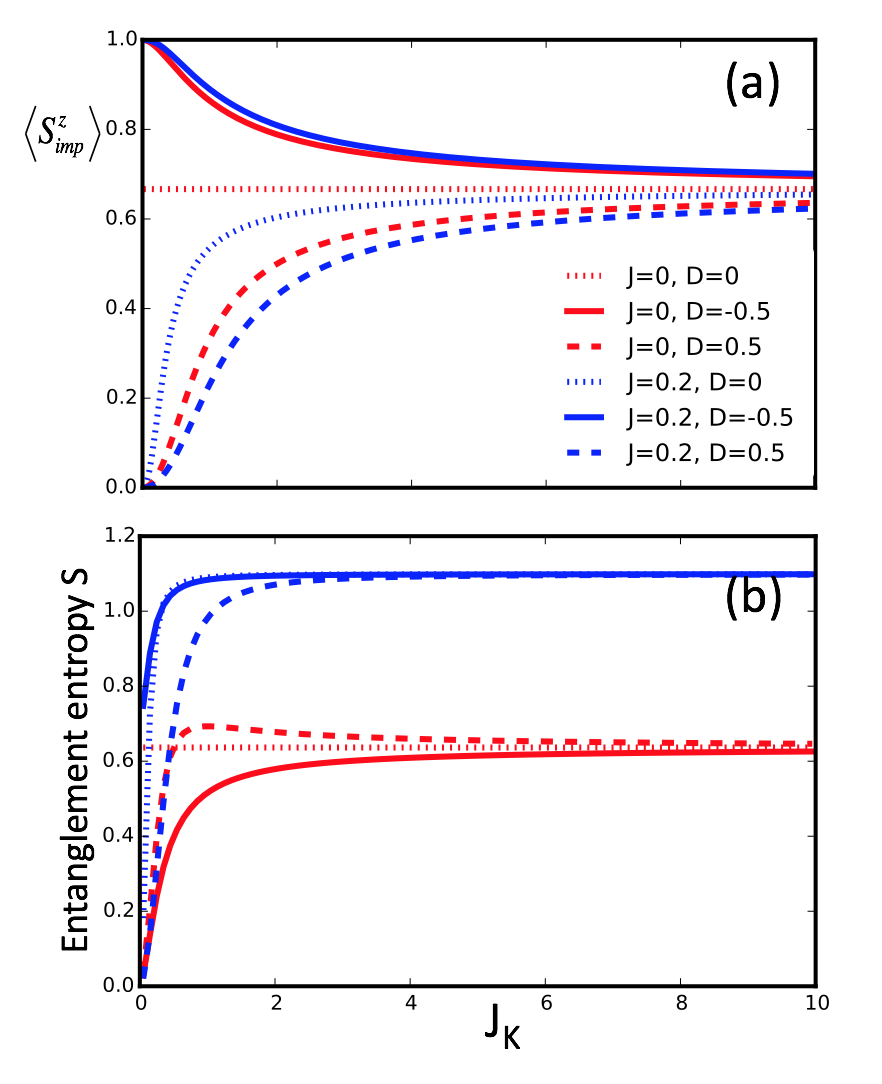}
\caption{(a) Impurity magnetic moment $\langle S^z_\mathrm{imp}
\rangle$ for the system of a single $S=1$ impurity connected to a
spin-$1/2$ via $J_K$, which in turn is connected to a second
spin-$1/2$ via $J$. We consider the ground state in the $S^z = 0$
sector. The antiferromagnetic interaction between the two
``conduction'' spins competes with the ``Kondo'' interaction. The
effect of $J$ is to induce a residual ferromagnetic coupling between
the second conduction spin and the impurity. This coupling tends to
reduce the impurity magnetic moment. In the strong coupling limit
$\langle S^z_{\rm imp}\rangle \rightarrow 2/3$. (b) Entanglement
entropy between the impurity spin and the two conduction spins.  Even
though $\langle S^z_\mathrm{imp}\rangle$ tends toward the same value
in the strong coupling limit, the entanglement entropy clearly
distinguishes two different regimes: for finite $J$ the impurity spin
is always fully screened, with an asymptotic value $S\rightarrow
\log{3}$. Without $J$, the impurity is always underscreened, with the
entropy converging to $S \rightarrow - \frac{1}{3} \log{(\frac{1}{3})}
- \frac{2}{3}\log{\frac{2}{3}}$. 
}
\label{fig:single_results}
\end{figure}

In order to account for this additional effect, we need to assume a residual interaction between the ``Underscreened'' states $|m_s=\pm1/2\rangle$ defined in Eq.~(\ref{ms}) and the rest of the Fermi sea. 
To make these ideas more concrete, we extend the toy model
with a second conduction electron spin coupled to the first by an anti-ferromagnetic exchange $J$, as depicted in Fig.~\ref{fig:single_model}(b).
We can anticipate that this interaction will counteract $J_K$, suppressing the magnitude of the magnetic moment of the impurity. More explicitly, we define a new basis: ${|0\uparrow\downarrow\rangle , |+\downarrow\downarrow\rangle, |0\downarrow\uparrow\rangle, |-\uparrow\uparrow\rangle}$. In this representation, the Hamiltonian is:
\begin{equation}
H=\begin{pmatrix}
-\frac{J}{4} & \frac{\sqrt{2}}{2}J_K & \frac{J}{2} & 0 \\
\frac{\sqrt{2}}{2}J_K & -\frac{J_K}{2}+\frac{J}{4}+D & 0 & 0 \\
\frac{J}{2} & 0 & -J/4 & \frac{\sqrt{2}}{2}J_K \\
0 & 0 & \frac{\sqrt{2}}{2}J_K & -\frac{J_K}{2}+\frac{J}{4}+D
\end{pmatrix}.
\label{h2}
\end{equation}
This matrix can be diagonalized to estimate the screening of the impurity moment $\langle S^z_{\rm imp} \rangle$. In Fig.~\ref{fig:single_results}(a) we show results for 
a fixed value $J=0.2$ and a range of magnetic anisotropies $D$ as a
function of $J_K$. 
For $J_K \lesssim J$, $D \neq 0$ the quantum fluctuations are greatly suppressed.
For $J_K \gtrsim J$, the values tends toward the value $S^z_{\rm
imp}=2/3$, as expected for the $|m_s=\pm1/2\rangle$ states.

\begin{figure}
\centering
\includegraphics[width=0.32\textwidth]{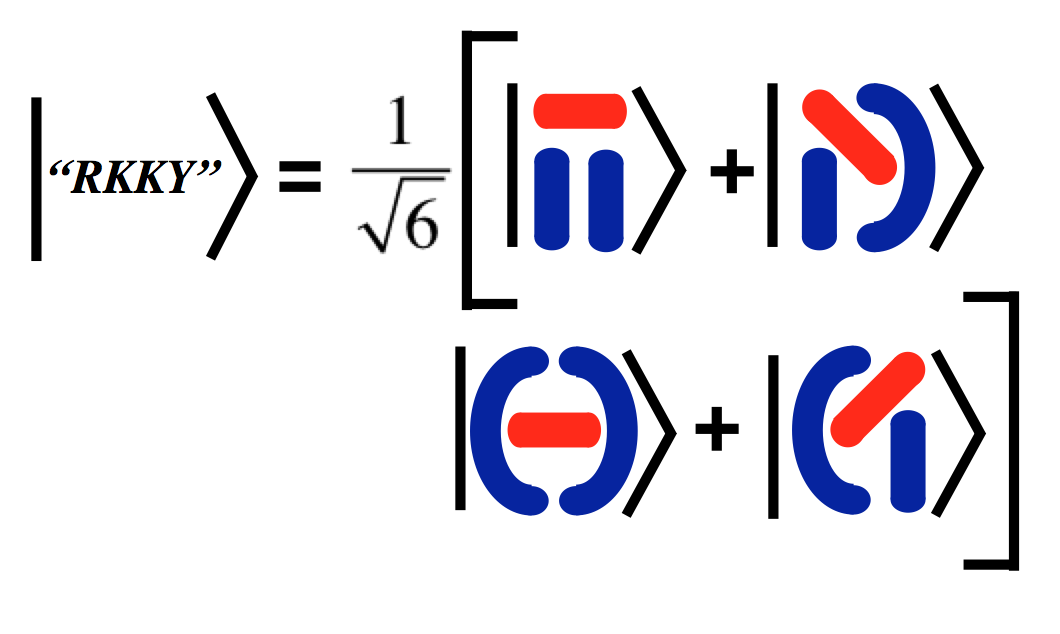}
\caption{Illustration of the RKKY state obtained for two impurities and two
``conduction spins''. Each impurity is coupled to a single conduction
electron via $J_K$, and the conduction electrons are coupled to each
other via antiferromagnetic $J$. The spin-1 impurity is represented as
two spins $S=1/2$. The dangling unscreened spins are shown explicitly
and the thick lines represent singlets between spins-1/2. In the
ferromagnetic case, the red dimers would be replaced by triplets.}
\label{fig:rkky}
\end{figure}

\subsubsection{Magnetic anisotropy effects}

It is instructive to observe the results for the anisotropic case in
Fig.~\ref{fig:single_results} more
closely. For
$D>0$, $J=0.2$, they look very similar to the isotropic case. In fact, finite $D$ also tends to force the impurity into the zero magnetization
state (for small $J_K$), enhancing the effects of $J$ and making the residual
ferromagnetism more effective. On the other hand, $D<0$ tends to
polarize the spin into the $S^z_{\rm imp}=1$ orientation. This time
the Kondo interaction has to compete against both, the anisotropy, and
the $J$ term, to reach the strong interaction value for large $J_K$.


For small $J$, one can use first-order degenerate perturbation theory
to show that the ground state will be fully screened, \textit{i.e.} $|{\rm
g.s.}\rangle = 1/\sqrt{2}\left(|m_s=1/2,\downarrow\rangle -
|m_s=-1/2,\uparrow\rangle \right)$. Here, the extra spin is entangled
with the dangling spin into a singlet, shown as the ``Screened'' state in Fig.~\ref{fig:screening}.
We point out that the actual ground state for $D \ll 0$ is in the $S^z=\pm 1$ sectors, with the impurity pointing in either direction, disentangled from the conduction spins, as illustrated by the states $|S=1;S^z=\pm 1 \rangle$ in Fig.~\ref{fig:screening}. 
These observations will become important in the analysis of the full
problem.

We now introduce a transverse anisotropy $E$ through the term (\ref{transverse}). 
If we ignore $J$, the net effect of $V_E$ is to mix $|m_s=1/2,\downarrow\rangle$ with $|-\downarrow\rangle$ and $|m_s=-1/2,\uparrow\rangle$ with $|+\uparrow\rangle$. The coupling $J$ to the third spin will yield a symmetric and antisymmetric linear combination of the two. 
The states $|- \downarrow\downarrow\rangle,|+ \uparrow\uparrow
\rangle$ have unperturbed diagonal energies $J_K/2+D$. If we assume $E
< J_K$ and $D \gg E$, this tunneling barrier is very high and the corrections to the ground state very small. 
Therefore, once again we conclude that this term does not change the
general picture in this parameter regime.

\begin{figure}
\centering
\includegraphics[width=0.48\textwidth]{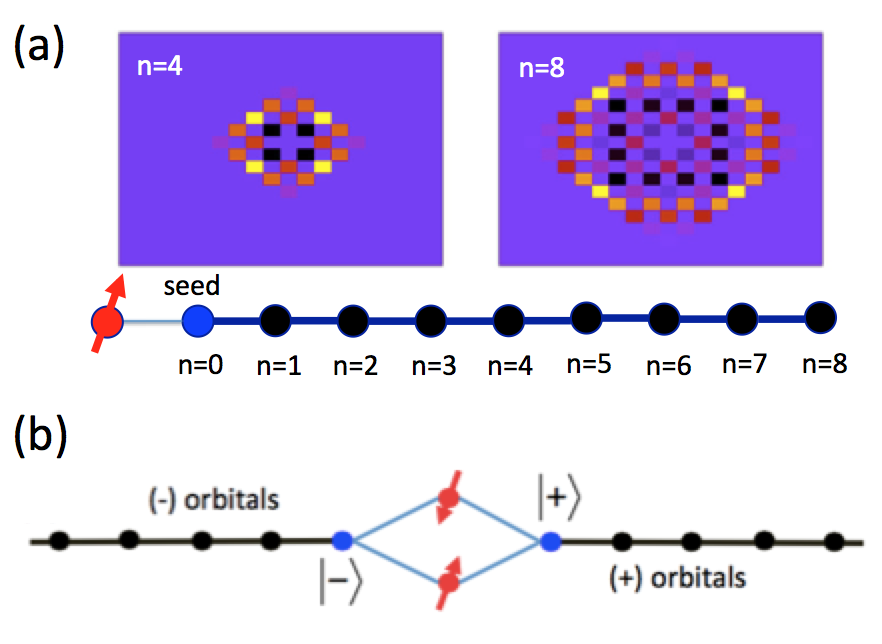}
\caption{(a) Chain representation for the single-impurity problem. The bulk non-interacting band structure is mapped via a canonical transformation onto a semi-infinite chain. The impurity is connected to the seed state. We also show the wave function amplitudes for the Lanczos orbitals $4$ and $8$ in real space. For the two-impurity problem (b), a folding transformation yields one-dimensional bonding and anti-bonding channels. The many-body terms couple the impurities to both. }
\label{fig:mapping}
\end{figure}

\subsection{Two impurities}\label{Two}

In order to provide intuition on the two-impurity problem, we consider
two single-impurity ``Underscreened'' states defined in Eq.~\eqref{ms}, and introduce a Heisenberg exchange between the conduction spins. This coupling can be either ferromagnetic or anti-ferromagnetic. Let us fix the total spin projection to $S^z=0$, in which case we have two degenerate states in the absence of exchange: $|1/2,-1/2\rangle$, and $|-1/2,1/2\rangle$, where the labels refer to the value of $m_s$. A simple analysis in terms of degenerate perturbation theory yields an effective Hamiltonian in the form of a $2\times2$ matrix that can be readily solved to yield that the ground state can be a singlet or a triplet, depending on the sign of the interaction:
\begin{equation}
|{\rm g.s.}\rangle =\frac{1}{\sqrt{2}}\left(|1/2,-1/2\rangle \pm |-1/2,1/2\rangle \right).
\label{gs_two}
\end{equation}
If we once again represent the spin-1 in terms of two spins-$1/2$'s, one
can find that for the isotropic case, this
corresponds to an equal superposition of all possible dimer coverings
between the dangling spins, as shown in Fig.~\ref{fig:rkky}. This
indicates that the RKKY state can be interpreted as the unscreened
spin becoming maximally entangled. The entanglement entropy between an
impurity spin an the rest of the system in this case is $S_{\rm RKKY}
= \log{3}$ and the impurity spin is fully screened.

From the ground state expression (\ref{gs_two}) we find that the spin-spin correlations are simply:
\begin{eqnarray}
\langle {\rm g.s.}|S_1^xS_2^x+S_1^yS_2^y|{\rm g.s.}\rangle & = & \pm 4a^2b^2 \nonumber \\
\langle {\rm g.s.}|S_1^zS_2^z|{\rm g.s.}\rangle & = & -b^4.
\label{corr_two}
\end{eqnarray}
In the isotropic case these values are $\pm 8/9$ and $4/9$, respectively. 
In the anisotropic case upon the inclusion of a finite value of $D$, the $|m_s\rangle$ wave functions will change in character from more Ising-like for $D < 0$, with $b > a$ in (\ref{ms}), to non-magnetic with $a > b$ for $D > 0$. Remarkably, the transverse correlations will reach a maximum value for $D=J_K/2$, since for this case $a=b=1/\sqrt{2}$, and only then they will start decreasing monotonically with increasing $D$. 

We now comment on the effects of the transverse anisotropy. Assuming that $J$ is the smallest energy scale in the problem, we should solve the single site problem first, and then introduce the RKKY coupling perturbatively. 
 Working in the subspace defined by 
 $|11\rangle,|22\rangle,|21\rangle,|12\rangle$, where the single impurity states $|1\rangle$ and $|2\rangle$ were defined in Eq.(\ref{states12}), we find that the
 interaction will only mix $|11\rangle$ with $|22\rangle$ and $|12\rangle$ with $|21\rangle$. If we focus on the last two, the Hamiltonian will be a bi-symmetric $2\times2$ matrix with eigenstates $1/\sqrt{2}\left(|12\rangle \pm |21\rangle \right)$, similar to what we had previously, but now with contributions from other spin sectors.

\begin{figure}
\centering
\includegraphics[width=0.44\textwidth]{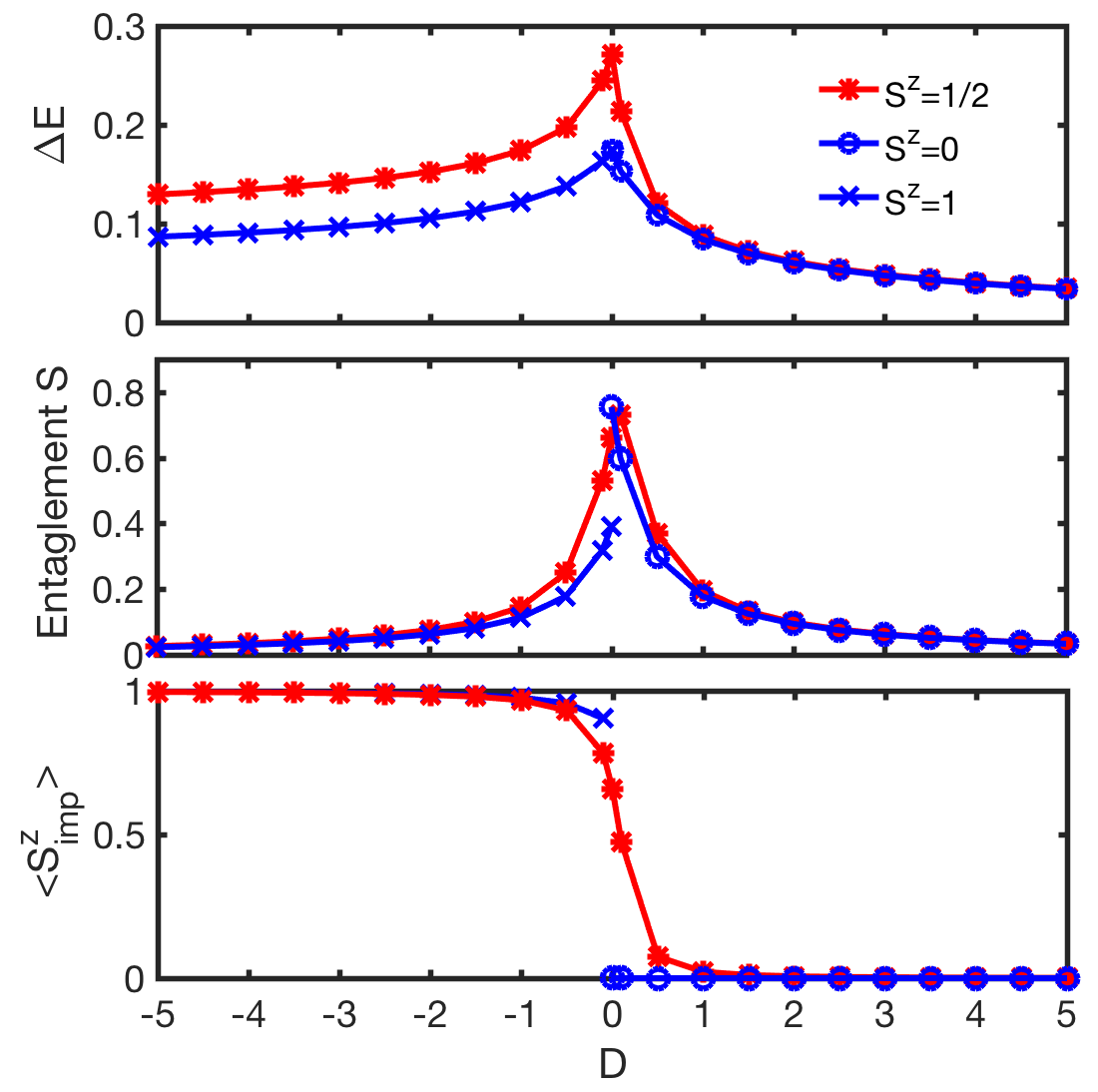}
\caption{(a) Energy gain as defined in Eq.~\eqref{egain} for the
single-impurity $S=1$ Kondo problem with $J_K$ = 1 for chain lengths 
$L=40$ in the $S^z = 0,1$ sectors, and $L=41$ in the $S^z = 1/2$ sector.
The $S^z=1/2$ sector corresponds to an underscreened impurity and is 
energetically favorable for finite systems.
(b) Impurity entanglement entropy and (c) expectation value of the
impurity moment, $\langle S^z_{\rm imp} \rangle$, 
for the same systems and parameters. These results show that for $|D|
\gtrsim J_K$ the physical quantities show only very weak even/odd
effects, but these are large for $|D| \lesssim J_K$.}
\label{fig:energy_entropy}
\end{figure}
\section{Methods}\label{Method}

\subsection{Single impurity}

In order to make the single-impurity problem amenable to DMRG
simulations, one needs to apply the unitary transformation described
in Ref.~\onlinecite{Busser2013}. The premise is to map the non-interacting
lattice Hamiltonian $H_{\rm band}$ in Eq.~(\ref{Hamiltonian}) onto a
non-interacting semi-chain, in the same spirit as Wilson's numerical
renormalization group method\cite{nrg_wilson,nrg_bulla}.  In the
absence of a lattice, Wilson chose a basis of partial waves that
expand radially away from the impurity.  We use a similar approach, in
which the basis is built recursively by following a number of very
simple steps (we refer the reader to the original
proposal\cite{Busser2013} for technical details): (i) we define a
single-particle seed orbital situated at the site connected to the
impurity, (ii) apply the non-interacting Hamiltonian to generate a
Krylov basis, and orthogonalize it following a Lanczos
procedure. The new single-particle wave functions
are orthogonal and connected to each other by a single
matrix element. The resulting Hamiltonian in the Lanczos basis will be
identical to a semi-infinite chain with site-dependent hoppings, see
Fig.~\ref{fig:mapping}(a).

We emphasize that this is an exact canonical transformation and that the
information about the lattice structure and the hybridization to the
impurity is completely preserved.  The many-body terms involve only
the first site of the chain (the seed orbital) and the impurity, and
remain local after the transformation. The resulting one-dimensional
problem can be easily solved with the DMRG method, with all numerical
errors under control. 

Our calculations are conducted on finite systems on the square
lattice. One of the remarkable aspects of the mapping is that the size
of the chain corresponds to the linear dimension of the original
system: a chain of length $L=50$ represents a square tilted 45
degrees with $~100$ sites along the diagonal. The ``missing''
orbitals in the Hilbert space live in different symmetry sectors that
do not couple to the impurity (similar to the NRG mapping, where only the $s$-wave sector is
preserved). The fact that the system is finite means that there is an
additional important energy scale, the level spacing $\Delta$, which
depends on $L$. For
values of the Kondo coupling $J_K < \Delta$ we would find the
impurities in the free moment regime, decoupled from the lattice. This
is not a problem in the work presented here, since all values of $J_K$
used are large enough.

\begin{figure}
\centering
\includegraphics[scale=0.6]{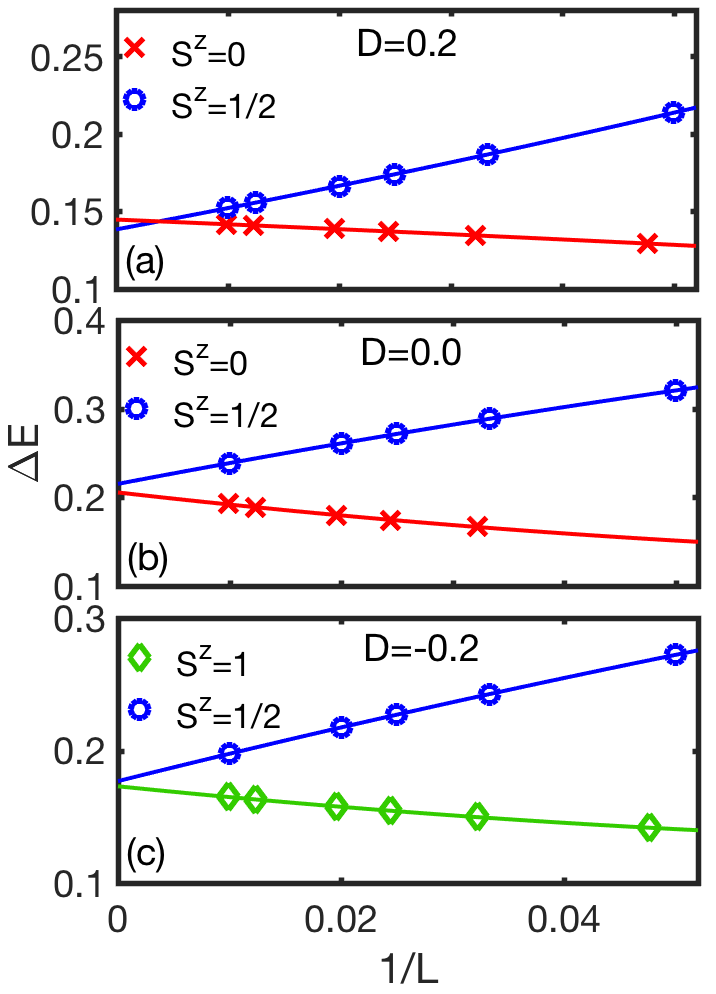}
\caption{Scaling of the energy gain $\Delta E$ with system size $1/L$ for $J_K=1$ and (a) $D=0.2$, (b) $D = 0$ and (c) $D = -0.2$. 
For finite systems the underscreened state in the $S^z=1/2$ sector
dominates in all three cases. Lines represent quadratic fit in $1/L$.
 Error bars in the extrapolation are of the order of the symbol size.}
\label{fig:finite_size}
\end{figure}

In addition, one has to take another important consideration into
 account: even and odd effects. In our present case, the correct
 length that will realize the actual ground state of the problem is
 not obvious {\it a priori}, so we resort to studying the energy
gained by connecting an impurity with $J_K$, compared to the energy of a disconnected impurity with $J_K=0$:
\begin{equation}
\Delta E(J_K) = E_0(J_K=0) - E_0(J_K).
\label{egain}
\end{equation}
The system will typically lower its energy by entangling the impurity to the conduction spins.

 The results for fixed $J_K=1$ as a function of anisotropy
 $D$ are shown in Fig.~\ref{fig:energy_entropy} and demonstrate that
 in finite systems odd chains with $S^z=1/2$ realize a ground state
 that is energetically favorable.  However, for $|D| \gtrsim J_K$ all
 physical quantities such as the impurity entanglement entropy and
 $\langle S^z_{\rm imp} \rangle$ are practically indistinguishable
 regardless of the chain lengths.

 As pointed out previously in Section II.A.2., the energy difference is dominated by the residual
 ferromagnetic interactions that are affected by the inter-level
 splitting in the non-interacting chains. This can be studied
 quantitatively by considering the energy gain as a function of
 systems size for even and odd chain lengths, see
 Fig.~\ref{fig:finite_size}. A quadratic fit and $L\to\infty$
 extrapolations indicate that the result in the thermodynamic limit is
 independent of the parity of the chain length, as expected. This is
 easy to understand: In a chain much longer than the Kondo length,
 flipping a spin or removing a site from the chain should make no
 difference in the physical behavior of the impurity. These effects
 have been elegantly explained in
 Refs.~\onlinecite{schwabe2012competition,Yang2017}, and are a direct
 consequence of working on finite systems. Odd-length chains present a
 state right at the Fermi level. When we turn on the Kondo interaction
 and this is smaller or of the order of the level splitting, first-order perturbation theory introduces a direct coupling between the
 impurity and this electronic state, which is solely responsible for
 the screening. This is why our description in Section ~\ref{Strong}
 correctly describes the physics. Even-length chains do not have a
 state at the Fermi level and the first-order correction vanishes. The
 energy gain due to the second-order correction is much smaller than
 that arising from the first-order term. Therefore, the differences
 seen in Fig.~\ref{fig:finite_size} are all a simple consequence of
 the inter-level spacing in finite chains.  These results underscore
 the importance of carefully handling the even-odd effects on finite systems and indicate
 that the proper route in finite systems is to take $S^z = 1/2$ for
 all values of $D$. 

\begin{figure}
\centering
\includegraphics[scale=0.5]{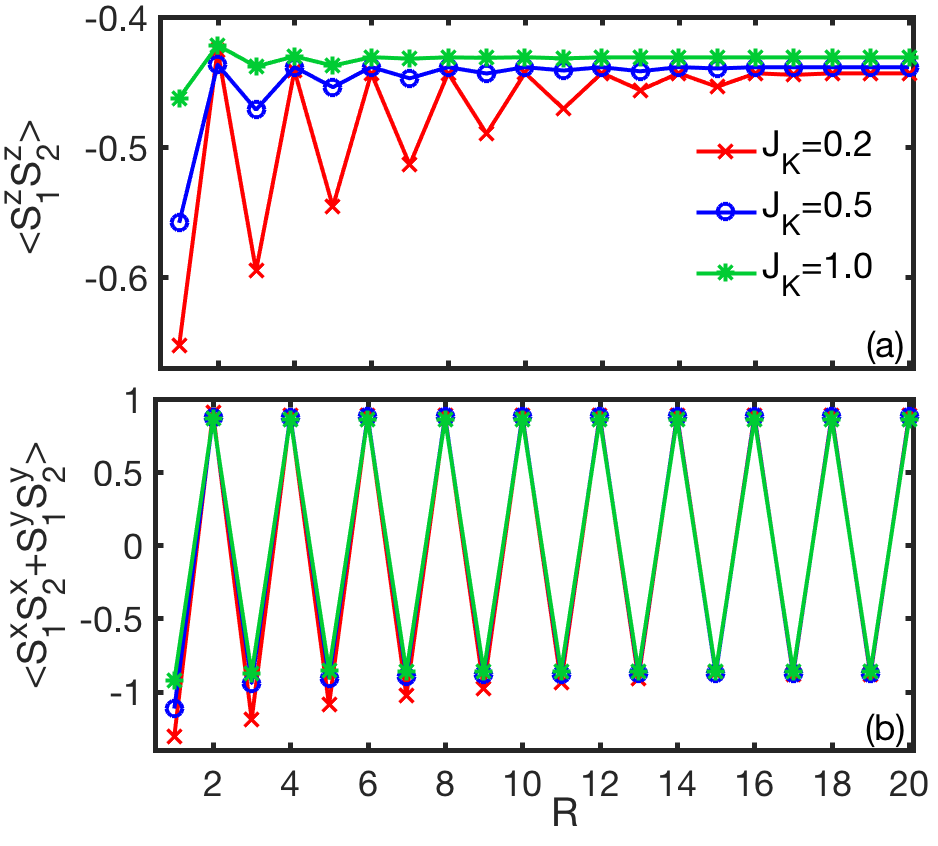}
\caption{Diagonal (a) and off-diagonal (b) spin-spin correlations 
between two spin-1 impurities as a function of their separation $R$
for a range of values of $J_K$ in the $S^z=0$ sector. We use two chains of length $L=61$ each. }
\label{iso1}
\end{figure}

\subsection{Two impurities}

In order to generalize the aforementioned procedure to the case of
multiple impurities, one applies a Block Lanczos transformation. The
resulting Hamiltonian is not block diagonal. In the particular case of
two impurities, this corresponds to a ladder-like geometry
\cite{Allerdt.Kondo,Allerdt.Graphene,Allerdt.Shockley,Allerdt.TI}. An
intuitive picture can be offered as follows: Suppose that we pick two
seed orbitals, situated at the positions of the two impurities. We can
apply the procedure outlined above for a single impurity: the orbitals
will start expanding away from each impurity, generating two
one-dimensional chains. At some point, when the length of the chains
is $~R/2$ ($R$ being the distance between impurities along the
lattice axis) the orbitals start overlapping and interfering,
translating into terms mixing the two chains, which remain
local.

We can, however, exploit the reflection symmetry with respect to the centre-point
between the two impurities by choosing the seed to be a bonding
(symmetric) or anti-bonding (anti-symmetric) linear combination of the
original local seeds. For each initial state, the Lanczos iteration
procedure is identical to that described in the single-impurity
problem\cite{Busser2013}, with two independent chains, one for each
channel, that do not mix. The resulting problem is then truly
one-dimensional, see Fig.~\ref{fig:mapping}(b). Under this
transformation, the many-body interactions are modified: there are
terms mixing the impurities and the first orbital of each chain. We
note that this symmetrization is identical in spirit and form to the
folding transformation used in the NRG calculations for the two-impurity
problem, \cite{Jones1989} with the main difference being that our
symmetrization takes place in real space instead of momentum space.
The advantage of this approach is not only that the recursion is
greatly simplified, but also that the equivalent one-dimensional
problem greatly reduces the entanglement, and thus the computational
cost of the simulations. 

We consider a square lattice at half filling, mapped so that 
each chain has an even number of sites. In the DMRG simulations the
truncation error was below $10^{-8}$, which implies
keeping up to 3600 states in some calculations. 

\begin{figure}
\includegraphics[scale=0.46]{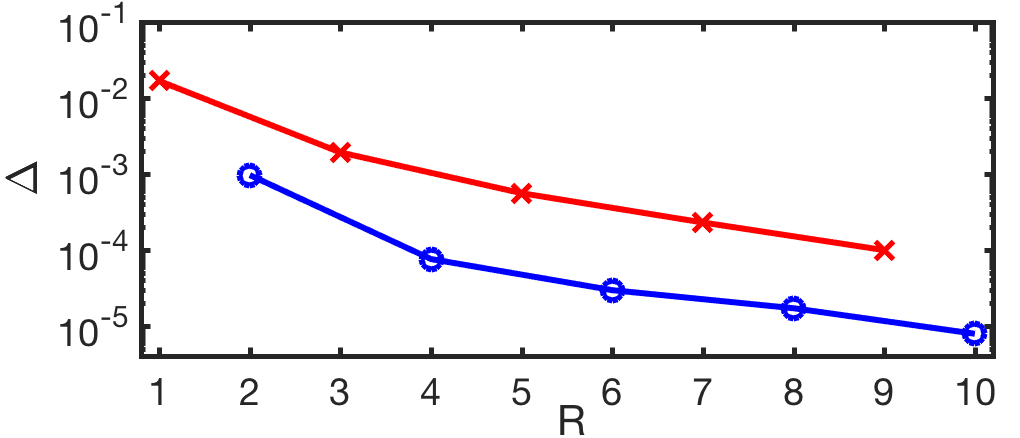}
\caption{(color online) Gap from the triplet (singlet) ground state to the first excited state at even distances (blue) and odd distances (red) for $J_K = 1.0, D=E=0, L=41$.}
\label{fig:even_gaps}
\end{figure}

\section{Results}\label{Results}

\subsection{Isotropic case: \quad $D=0, E=0$}

In this section we present the results for the spin-1 two-impurity problem in the
absence of anisotropy. Based on the intuition established through
the studies of the spin-$1/2$ two-impurity problem, two types of
behaviors could be expected: formation of two independent Kondo
states, or coupling  via
indirect exchange, with both processes in competition dictated by the relative
positions of the two spins and the magnitude of the Kondo interactions
$J_K$. We will show that the spin-$1$ case is more complex.

In Fig.~\ref{iso1}, the spin-spin correlations $\langle S^z_1
S^z_2\rangle$ between impurities are shown in the $S^z=0$ sector as a
function of the distance between the impurities, $R$, for different values
of $J_K$. The spins are positioned along the $x$ axis, therefore even
(odd) distances correspond to impurities on the same (opposite)
sublattice.  Since the RKKY interaction has different character in
these two cases, we discuss them separately for clarity.

\begin{figure}
\includegraphics[scale=0.49]{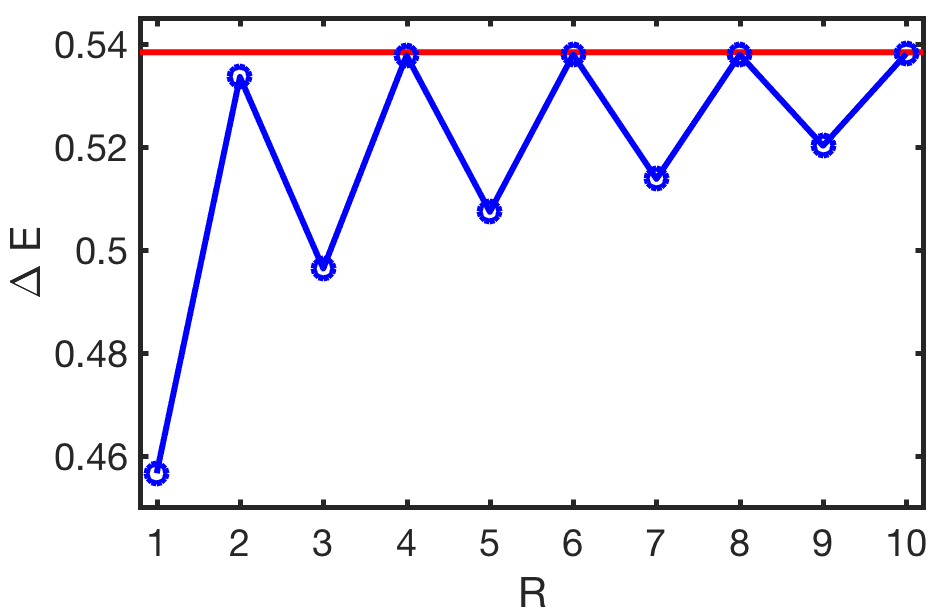}
\caption{Energy gain for two $S=1$ impurities as a function of distance, for $J_K=1, L=41$. The horizontal line represents the value for two independent impurities at infinite distance. }
\label{fig:egain2}
\end{figure}

\subsubsection{Opposite sublattices ($R=1,3,\cdots$)}

When impurities are on opposite sublattices, the ground state is a
 singlet, $S=0$. As seen in the lower panel of Fig.~\ref{iso1}, the
 magnitude of transverse correlations is twice the value of
 diagonal correlations, which is a signature of a singlet state. With
 increasing $J_K$, the correlations approach algebraically the
 limiting value of $\langle S^z_1S^z_2 \rangle= -(\frac{2}{3})^2 =
 -4/9$. This result is explained by considering two independent $S=1$
 problems as discussed in Section~\ref{Single}: The case $S^z=0$, $J_K
 \rightarrow \infty$ can be described as one impurity forming an
 $m_s=1/2$ state with a conduction electron, while the other forms an
 $m_s=-1/2$ state with another conduction electron, resulting in a four-fold degenerate state. From the explicit form of
 the wave-functions, calculated in Section~\ref{Single}, it follows that
 the correlations should equal the stated value. The same value is
 also reached in the large-$R$ limit for any finite value of $J_K$.
 
 These results show that the dangling spins (residual spin-$1/2$ local
 moments) couple anti-ferromagnetically through the indirect RKKY
 exchange. This means that, unlike in the case of spin-1/2 Kondo
 impurities, {\it Kondo and RKKY physics coexist}.

\subsubsection{Same sublattice ($R=2,4,\cdots$)}

Impurities on the same sublattice yield a triplet ground state. However, the gap to the excited non-degenerate singlet state is quite small. 
Calculations were done with high energy precision and varying system
sizes, $L=21$, 41 and 61 (where $L$ is the size of each
non-interacting chain without the impurity), confirming that this gap
is not a numerical artifact. The splitting
decreases with increasing impurity separation, as seen in Fig.~\ref{fig:even_gaps}, and barely changes with chain size. The system is gapless in the thermodynamic limit: flipping the spin of an electron very far from the impurity should not affect the physics. But this finite-size gap is dictated by the level splitting in the bulk, which is much larger than the singlet-triplet gap measured in this plot. 
We interpret this result as follows: the two impurities form a triplet (or singlet, depending on the sublattice) state mediated by the RKKY interaction. In a similar fashion as Nozi\`eres's Fermi liquid picture for the single impurity Kondo problem, this ``bound state'' acts as a scattering center for the conduction electrons that form an orthogonal Fermi sea (notice that in the chain representation, the two impurities are localized). The internal structure of the RKKY state is as depicted by Fig.~\ref{fig:rkky} and Eq.~\ref{gs_two}. From our discussion in Section \ref{Two} we learn that the singlet-triplet gap is dictated by the residual interactions with the conduction electrons that are responsible for mediating the RKKY exchange.

\begin{figure}
\centering
\includegraphics[scale=0.5]{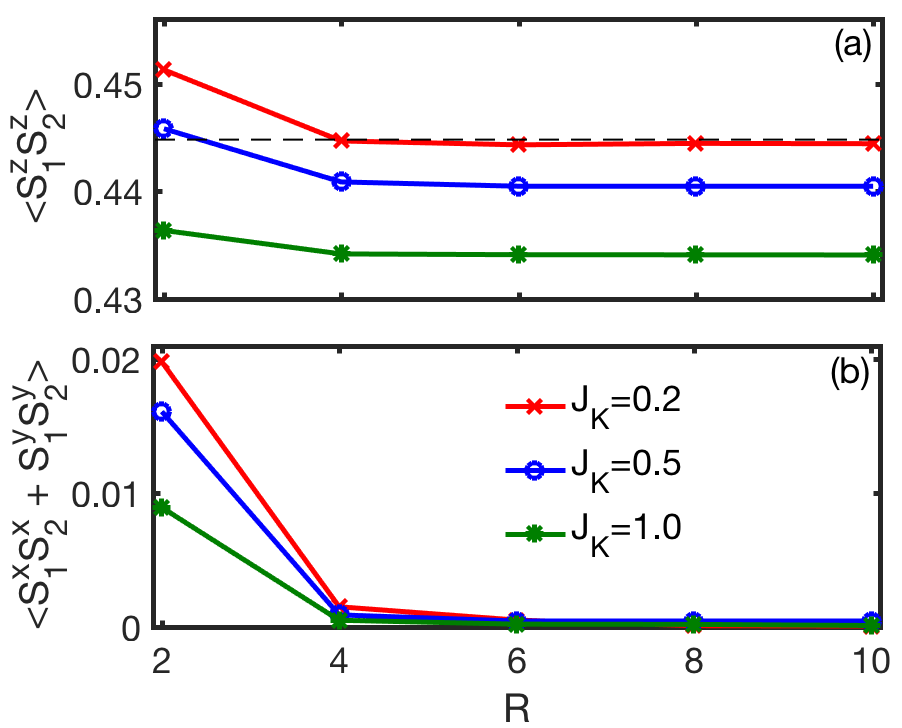}
\caption{Spin-spin correlations between two spin-1 impurities as a
function of their separation for a range of values of $J_K$ in the $S^z=1$ sector. 
The horizontal dashed line indicates the value of $4/9$. We only show even sites, corresponding to a triplet ground state. $L=61$. }
\label{iso2}
\end{figure}

Fig.~\ref{fig:egain2} shows the
energy gain for two impurities obtained using Eq.(\ref{egain}) with
$J_K=1$. This quantity provides information about the correlation energy. We notice that for impurities on the same sublattice and for sufficiently large $R$, $\Delta E$ approaches a constant value: twice the energy gain for a single impurity. This indicates that they tend to form two independent Kondo clouds, oblivious to the presence of the other spin and  with the consequent four-fold degeneracy due to the two dangling spin 1/2's pointing in arbitrary directions.
This occurs already at relatively short distances and is a remarkable aspect of the two impurity problem on a half-filled bipartite lattice: the electronic wave functions have nodes at the position of the second impurity hindering the possibility of mediating an indirect RKKY exchange\cite{Allerdt.Kondo}.
The energy gain, or correlation energy (Fig.~\ref{fig:egain2}) combines the contributions from the partial Kondo screening (proportional to $J_K$) and the RKKY interaction. Therefore, it is not the energy gain, but the finite-size gap (Fig.~\ref{fig:even_gaps}) that yields a measure of $J_{RKKY}$.

\begin{figure}
\centering \includegraphics[scale=0.5]{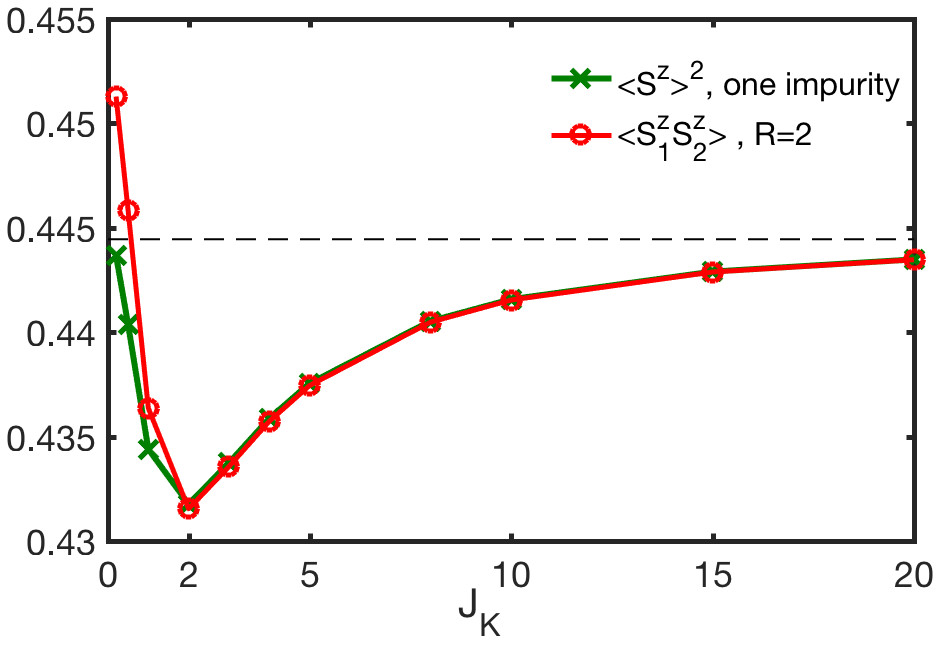}
\caption{Spin-correlation between two impurities at fixed separation
$R=2$ as a function of the Kondo coupling $J_K$. The single-impurity moment squared,
$\langle S^z \rangle^2$, is also shown for comparison. For $J_K > 2$,
the two lines are almost indistinguishable, indicating two
essentially uncorrelated
underscreened impurities. The chain lengths are $L=84$ and $L=40$ for
the two-impurity and single-impurity cases, respectively.} 
\label{iso3}
\end{figure}

Rigorously, the strict ground state at even distances is perfectly
described in terms of two entangled Kondo states forming a triplet, as
obtained in Section \ref{Two}. Due to the choice of $S^z=0$ in the
DMRG calculations, we cannot observe ferromagnetic alignment in the
diagonal correlations, however the off-diagonal correlations are {\it
minus} twice the value of the diagonal ones, which is a signature of a
triplet state (see Fig.~\ref{iso1}).  Since the system is degenerate, we can also consider
the ground-state in the $S^z=1$ sector (Fig.~\ref{iso2}) to confirm this interpretation. In this case, the dangling spins point in the same
direction and we expect the correlations in the $z$-direction
to converge to $b^4=(\frac{2}{3})^2$, which is indeed observed in
Fig.~\ref{iso2}.

Interestingly, the correlations only saturate to $(\frac{2}{3})^2$ in
the strong-coupling limit and have a non-monotonic behavior as a
function of $J_K$. This is highlighted in Fig.~\ref{iso3} for distance $R=2$.  The reason the correlations dip below
$(\frac{2}{3})^2$ is the residual ferromagnetic coupling of the spin-1
Kondo effect. This is a well-known single-impurity phenomenon. At weak
coupling, the residual ferromagnetic interaction with the conduction
electrons reduces the absolute value of the
correlations\cite{Nozieres1980}. Upon increasing $J_K$, the system
tends toward the strong coupling limit and the residual ferromagnetism
disappears. However, the impurities completely disentangle before that
limit is reached: at values of $J_K > 2$ the spin-spin correlations
are $\langle S^z_1 S^z_2 \rangle = \langle S^z_{\rm imp} \rangle ^2$,
where $\langle S^z_{\rm imp} \rangle$ is the result for a single
impurity. The same effect would be seen for odd distances, with the difference being that the correlations are negative.



\begin{figure}
\centering
\includegraphics[scale=0.4]{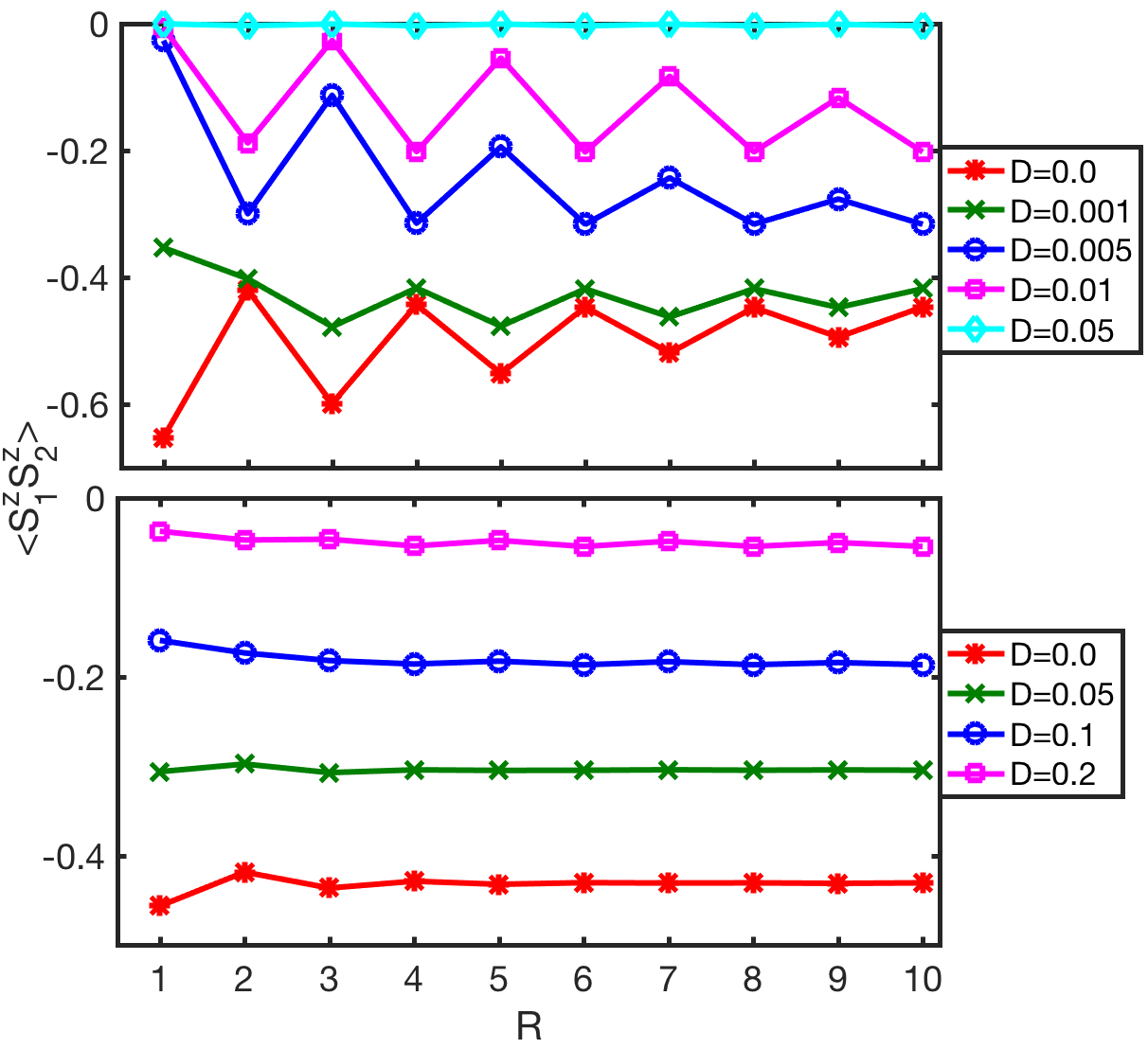}
\caption{Spin-correlations between two spin-1 impurities as a function
of their separation $R$ for different values of longitudinal magnetic
anisotropy $D>0$ with $J_K=0.2$ (top) and $J_K=1.0$ (bottom). Here $L=104$.}
\label{aniso1_new}
\end{figure}

\subsection{Anisotropic case: \quad $D\neq0$~,~$E=0$}

Once magnetic anisotropy is included (Eq.~\ref{anisotropy}), the
impurities display different behavior depending on the sign of $D$.
For clarity we discuss the two cases separately.

\subsubsection{$D>0$}

When the anisotropy is positive, the ground state has $S^z=0$ regardless of what sublattice the impurities are coupled to. Fig~.\ref{aniso1_new} shows the $z$-component of the correlations for a small ($J_K = 0.2$) and large ($J_K = 1.0$) coupling.
For positive $D$, the impurity spin gets suppressed and the correlations shift monotonically to zero.
We do not observe ferromagnetism in the longitudinal direction and the
correlations remain always antiferromagnetic, irrespective of distance.

In all cases, the correlations at long distances
plateau at a constant value $\langle S^z_1 S^z_2 \rangle \rightarrow
-b^4$ that depends on $J_K$ and $D$. 
Moreover, for impurities on the
same sublattice we find that the ground state and the first excited
state are quasi degenerate, with a gap of the order of $10^{-5}$ or
smaller, indicating that the two dangling spins are practically
uncorrelated. As discussed in the isotropic case, the dangling spins
may point in either direction, leading to a four-fold (quasi)
degeneracy.

\begin{figure}
\centering
\includegraphics[scale=0.41]{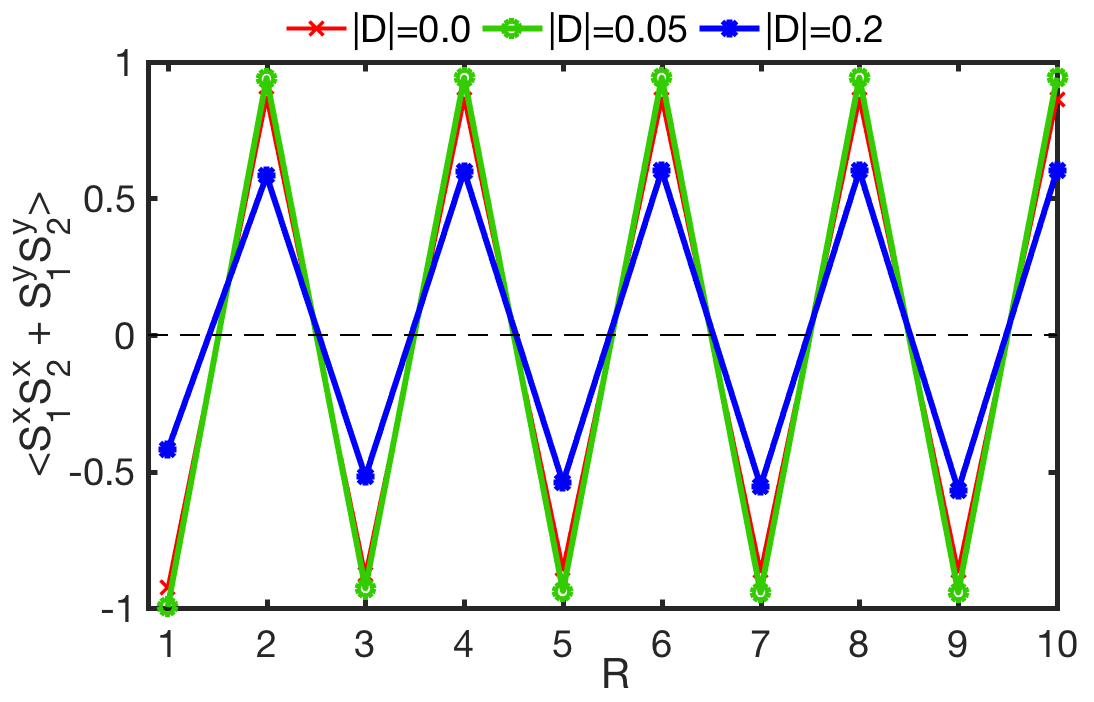}
\caption{Spin-correlations in the transverse direction for $J_K=1.0$ with $D>0$.}
\label{sxsx_anis}
\end{figure}

Due to the anisotropy, the correlations have different behavior in the
transverse direction, as seen in Fig.~\ref{sxsx_anis}. We first notice
that for small values of $D>0$ the magnitude of the correlations {\it
increases}. This was already observed in Sec.\ref{Two} for the simple
case of two sites. For large $|D|$, the correlations decay to zero,
indicating the suppression of the quantum fluctuations. In addition,
we can see the character of the RKKY interaction, oscillating between
ferromagnetic and antiferromagnetic depending on the sublattice.

\begin{figure}
\centering
\includegraphics[scale=0.42]{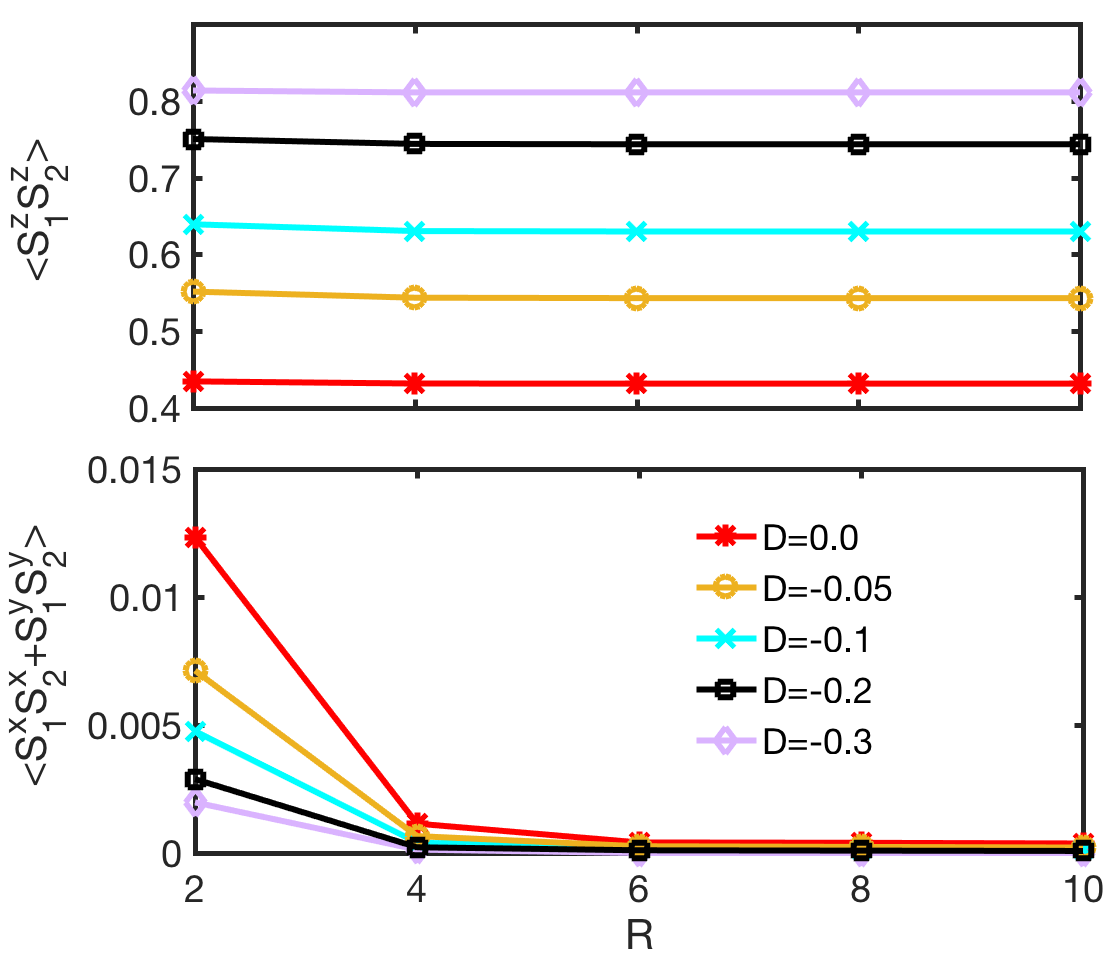}
\caption{Longitudinal (top) and transverse (bottom) spin-correlations for $J_K=1.0$ with $D<0$ at even distances for $S^z=1$.}
\label{aniso3}
\end{figure}

\subsubsection{$D<0$}

\begin{figure}
\centering
\includegraphics[scale=0.46]{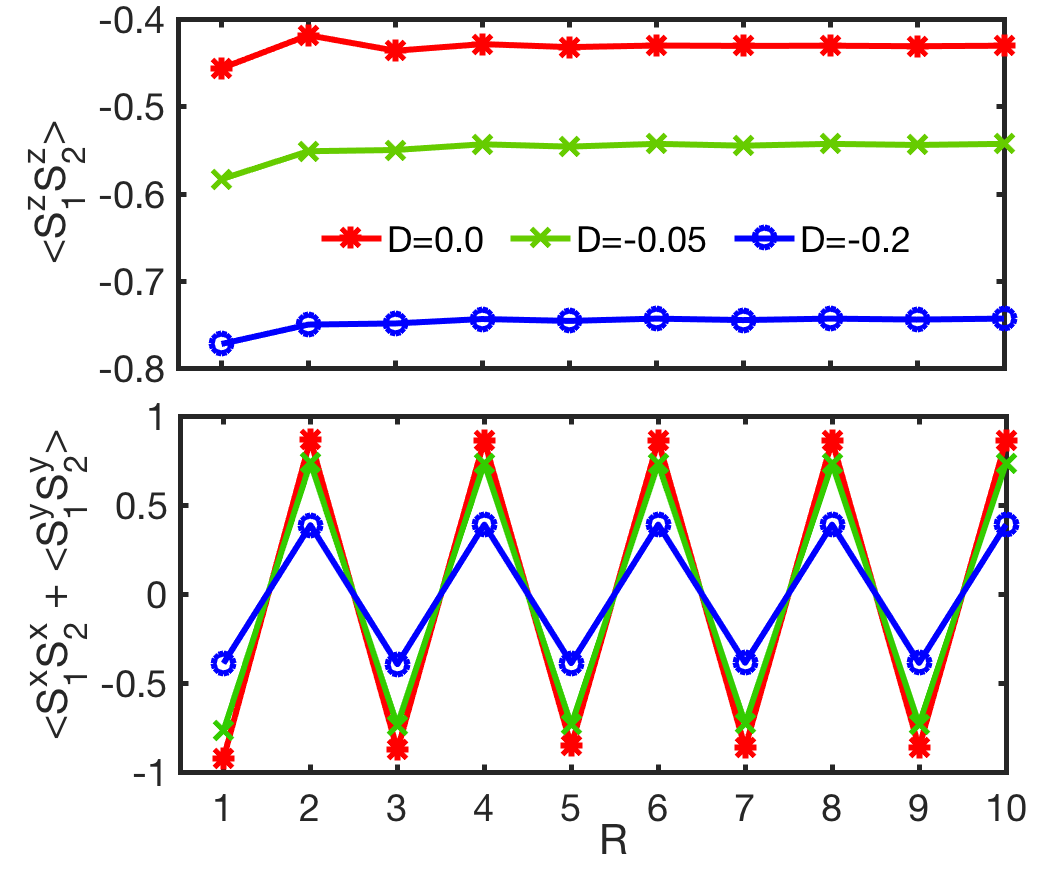}
\caption{Longitudinal (top) and transverse (bottom) spin-correlations for $J_K=1.0$ with $D<0$ with $S^z=0$.}
\label{aniso4}
\end{figure}

We switch now to the case of $D<0$. In this situation, the ground state has $S^z=0$ if impurities are on opposite sublattices, and $S^z=\pm1$ when on the same sublattice. Results for even distances are shown in Fig.~\ref{aniso3}. Here, the impurities tend to align with increasing $D$, approaching the values $\langle S^z_1S^z_2 \rangle = 1$ and $\langle S^x_1S^x_2 \rangle = 0$, as expected from Ising spins. The same argument applies to odd distances. In this case, the impurities are anti-aligned but still behaving as Ising spins as $D\rightarrow -\infty$, as seen in Fig.~\ref{aniso4}. Results for even distances in the $S^z=0$ sector are shown for comparison. Again these states could be described as \textit{quasi}-degenerate, with the gap on the order of $10^{-5}$.

\subsubsection{Characterization through entanglement}

\begin{figure}
\centering
\includegraphics[scale=0.5]{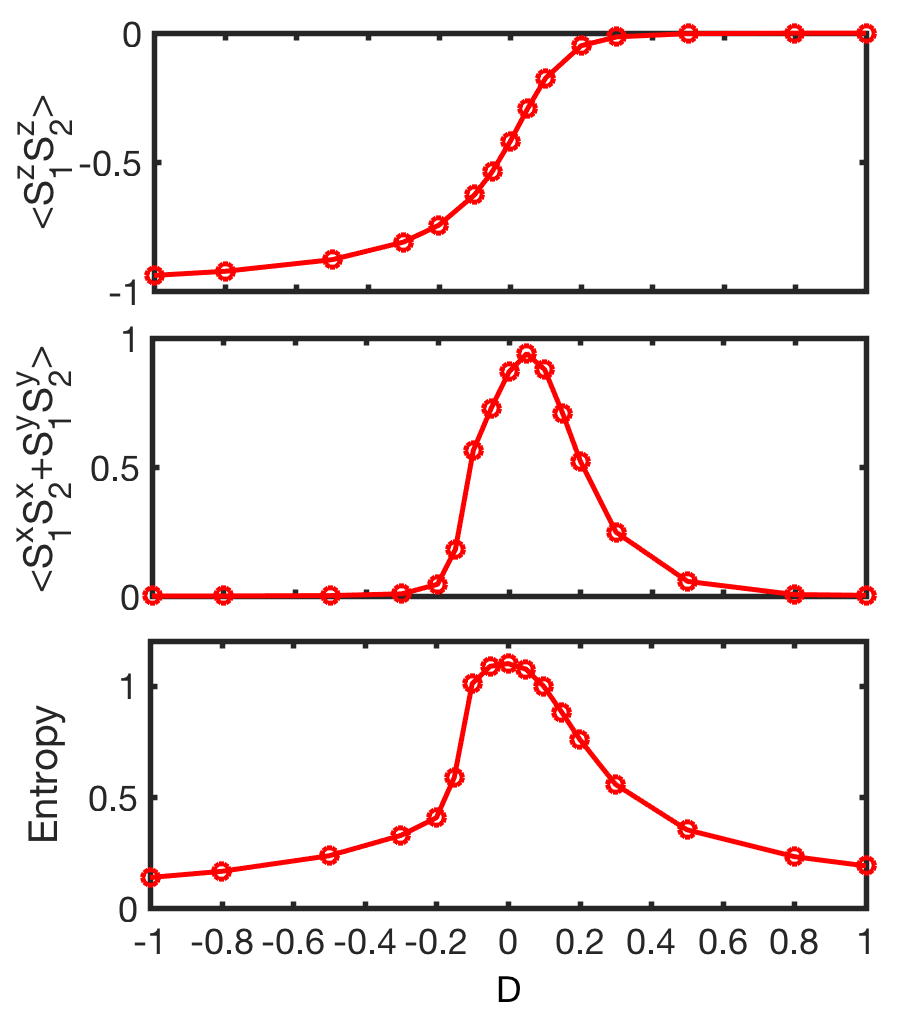}
\caption{(a) Diagonal and (b) off-diagonal spin-spin correlations
between impurities at distance $R=2$ for $J_K=1$ as a function of magnetic anisotropy $D$ and for $S^z=0$. We show results for chains of length $L=2n+1$. (c) Entanglement entropy for one of the impurities, obtained by tracing over the rest of the system. }
\label{fig:entropy_and_correlations}
\end{figure}

The screening process can be characterized through the entanglement
between one impurity spin and the rest of the system, shown in
Fig.~\ref{fig:entropy_and_correlations}, together with the spin
correlations. We considered the impurities at distance $R=2$ and
$J_K=1$, with chains of size $L=51$. Magnetic anisotropy (finite
$D$) tends to kill the spin fluctuations. For sufficiently large $D<0$
the impurities behave as Ising spins. For $D > 0$ the
impurities become practically uncorrelated in the $z$-direction, but
antiferromagnetic correlations survive in the transverse direction. This is consistent with the renormalization group
analysis\cite{Zitko.Properties_of_magnetic_impurities} that indicates
that the system should develop a dominant transverse anisotropy for $D> 0$. 

\subsection{Transverse anisotropy: \quad $D < 0$~,~$E \neq 0$}

\begin{figure}
\centering
\includegraphics[scale=0.42]{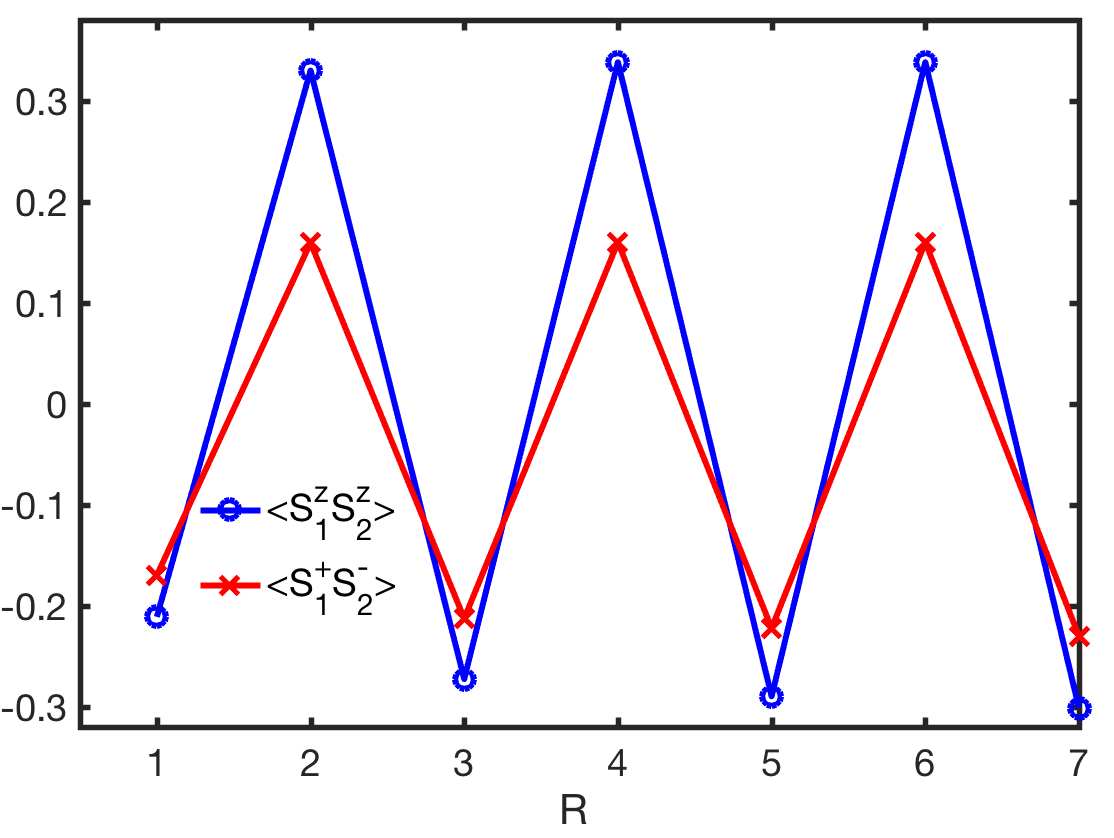}
\caption{Diagonal and transverse spin-spin correlations for impurities at different distances and parameters $J_K=1.0, D=-0.2, E=0.05$.}
\label{fig:transverse}
\end{figure}

Including the transverse anisotropy term $E$, see
Eq.~(\ref{transverse}), mixes subspaces with different
total $S^z$, making calculations considerably costlier. We looked at
one particular case, for parameters $ J_K=1.0, D=-0.2, E=0.05$ and
calculated the spin-spin correlations as a function of the
inter-impurity distance, shown in Fig.~\ref{fig:transverse}.  In this
case, the ground state is always non-degenerate but the gap still remains
very small when
the spins are on the same sublattice and decreases with the
inter-impurity distance. This indicates that for even
distances we still have an $S=1/2$ degree of freedom that practically
remains dangling. The effects of the transverse anisotropy are quite
dramatic, when comparing to Figs.~\ref{aniso1_new} and \ref{sxsx_anis}: The transverse
correlations, even though they preserve the same structure, are
considerably reduced, while the longitudinal one now present the
character proper of the RKKY interaction and oscillate between
ferro and anti-ferromagnetic.  This oscillation is now observable because the
interaction is mixing states with impurities pointing in either
direction. 

\section{Conclusions}

We have conducted an extensive study of the two-impurity Kondo problem
for spin-1 adatoms on square lattice. We provided a simple intuitive
picture and identified the different regimes, depending on system
size, Kondo coupling $J_K$ and magnetic anisotropy.
For two impurities, the nature and properties of the ground state 
depend most importantly on the spins being on the same or opposite
sublattices. Impurities on opposite (same) sublattice have a singlet
(triplet) ground state. However, the energy difference between the
triplet ground state and the singlet excited state is very small and
we expect a four-fold degenerate ground state for impurities on the
same sublattice. For large enough $J_K$ the impurities become
completely uncorrelated forming two independent underscreened states
with the conduction electrons.

Interestingly, our calculations support a picture in which underscreening (Kondo) and RKKY correlations coexist: conduction electron partially screen the individual impurities, and the dangling $S=1/2$ degree of freedoms are responsible for establishing RKKY correlations between them.

In order for the impurities to realize quasi-classical behavior we
need to go to the large $D < 0$ limit. For anisotropies of the order
of the Kondo coupling the impurities always display an important
amount of entanglement. The dependence of the entanglement and
correlations with system size indicate that this behavior could be
realized in experiments by electrostatically confining the impurity to an ``electron puddle'' (Kondo box)  and varying the size of the confining potential
around the impurity. 


\acknowledgements
We thank C. Hirjibejedin for useful and stimulating conversations.
AEF and AA are greateful to the U.S. Department of Energy, Office of Basic Energy Sciences, for support under grant DE-SC0014407.
R\v{Z} acknowledges the support of the Slovenian Research Agency (ARRS) under P1-0044 and J1-7259.


\end{document}